\documentclass[eqsecnum,preprint,prd,aps,nofootinbib]{revtex4}
\usepackage{epsfig}
\usepackage{amssymb}

\begin{document}
\begin{center}
\bibliographystyle{article}
{\Large \textsc{A pedagogical introduction to the Slavnov formulation
of quantum Yang--Mills theory}}
\end{center}

\date{\today}

\author{Hossein Ghorbani}
\email{ghorbani@to.infn.it}
\affiliation{Universit\`a degli Studi di Torino, 
Dipartimento di Fisica Teorica,
Via Pietro Giuria 1, 10125 Torino, Italy}
\affiliation{Istituto Nazionale di Fisica Nucleare, Sezione di
Torino, Via Pietro Giuria 1, 10125 Torino, Italy}
\author{Giampiero Esposito}
\email{giampiero.esposito@na.infn.it} 
\affiliation{Istituto Nazionale di
Fisica Nucleare, Sezione di Napoli, Complesso Universitario di
Monte S. Angelo, Via Cintia Edificio 6, 80126 Napoli, Italy}
\begin{abstract}
Over the last few years, Slavnov has proposed a formulation of
quantum Yang--Mills theory in the Coulomb gauge 
which preserves simultaneously manifest
Lorentz invariance and gauge invariance of the ghost field Lagrangian.
This paper presents in detail some of the necessary calculations, i.e.
those dealing with the functional integral for the S-matrix and its
invariance under shifted gauge transformations.
The extension of this formalism to quantum gravity in the Prentki
gauge deserves consideration. 
\end{abstract}

\maketitle

\section{Introduction}

Within the framework of a global approach to quantum theory 
\cite{dewi03}, the path integrals of ordinary 
quantum mechanics \cite{feyn48} and
the functional integrals of quantum field theory remain a tool of
fundamental importance and well suited for a Lagrangian, and hence
fully relativistic, quantization. In the sixties, the work of Refs.
\cite{dewi67a,dewi67b,fadd67} led to a unified description of quantum
Yang--Mills and quantum general relativity, and deeper foundations
were developed along the years until recent times \cite{dewi03}. 
More recently, new formal developments were obtained in Ref. 
\cite{slav08}, i.e.
\vskip 0.3cm
\noindent
(i) A formulation of quantum Yang--Mills theory which is manifestly
Lorentz invariant and leads to gauge invariance of the ghost-field
Lagrangian.
\vskip 0.3cm
\noindent
(ii) The problem of Gribov copies \cite{grib78} is avoided.
\vskip 0.3cm
\noindent
(iii) Perturbative renormalization still holds \cite{quad10},
despite the occurrence of a propagator that does not decrease at
infinity sufficiently fast \cite{slav08}.

The starting point of the analysis in Ref. \cite{slav08} is the
functional-integral representation of the S-matrix in the Coulomb gauge
for an $SU(2)$ gauge model, i.e.
\begin{equation}
S=\int {\rm exp} \left \{ i \int \Bigr[L_{YM}
+\lambda^{a} \partial_{j} A_{j}^{a} \Bigr] dx \right \} d\mu,
\label{(1.1)}
\end{equation}
where the measure $d\mu$ includes the Faddeev--Popov determinant 
\cite{fadd67}, and $L_{YM}$ is the standard Yang--Mills Lagrangian 
\begin{equation}
L_{YM}=-{1\over 4}F_{\; \mu \nu}^{a} \; F_{\; \mu \nu}^{a},
\label{(1.2)}
\end{equation}
built from the field strength
\begin{equation}
F_{\; \mu \nu}^{a}=\partial_{\mu}A_{\nu}^{a}
-\partial_{\nu}A_{\mu}^{a} +g \varepsilon^{abc}A_{\mu}^{b}A_{\nu}^{c}.
\label{(1.3)}
\end{equation}
Unlike the Abelian case, gauge invariance of the action in the integral
(1.1) is broken not only by the gauge-fixing term but also by the
Faddeev--Popov \cite{fadd67} ghost Lagrangian. To avoid this, the author
of Ref. \cite{slav08} has proposed consideration of the 
functional integral
\begin{eqnarray}
S &=& \int {\rm exp} \left \{ i \int \Bigr[L_{YM}
+(D_{\mu}\varphi^{*}) (D_{\mu}\varphi)
-(D_{\mu}\chi^{*})(D_{\mu}\chi) \right . \nonumber \\
&+& \left . (D_{\mu}b^{*}) (D_{\mu}e)
+(D_{\mu}e^{*})(D_{\mu}b)\Bigr]dx \right \}
\delta(\partial_{j}A_{j})d\gamma,
\label{(1.4)}
\end{eqnarray}
where the (formal) measure $d\gamma$ differs from $d\mu$ in (1.1)
by the product of differentials of the scalar fields
$$
(\varphi,\varphi^{*},\chi,\chi^{*},b,b^{*},e,e^{*}),
$$
where the fields $\varphi,\chi$ are commuting while $b,e$ are
anticommuting.

On making in the integral (1.4) the shift of integration variables
\begin{equation}
\varphi \rightarrow \varphi+{{\hat a}\over g},
\label{(1.5)}
\end{equation}
\begin{equation}
\chi \rightarrow \chi -{{\hat a}\over g},
\label{(1.6)}
\end{equation}
where
\begin{equation}
{\hat a} \equiv \left(0,{a \over \sqrt{2}}\right),
\label{(1.7)}
\end{equation}
$a$ being a constant parameter, the formula (1.4) yields
\begin{eqnarray}
{\widetilde S} &=& \int {\rm exp} \left \{ i \int \Bigr[
L_{YM}+(D_{\mu}\varphi^{*})(D_{\mu}\varphi)
+{1\over g}(D_{\mu}\varphi^{*})(D_{\mu}{\hat a}) \right . 
\nonumber \\
&+& {1\over g}(D_{\mu}{\hat a}^{*})(D_{\mu}\varphi)
+{1\over g}(D_{\mu}\chi^{*})(D_{\mu}{\hat a})
+{1\over g}(D_{\mu}{\hat a}^{*})(D_{\mu}\chi) 
\nonumber \\
&-& \left . (D_{\mu}\chi^{*})(D_{\mu}\chi)
+(D_{\mu}b^{*})(D_{\mu}e)
+(D_{\mu}e^{*})(D_{\mu}b)\Bigr]dx \right \}
\delta(\partial_{j}A_{j})d\gamma.
\label{(1.8)}
\end{eqnarray}
Interestingly, using different types of scalars for the kinetic
terms, $(\varphi,\chi)$ being commuting and $(b,e)$ being anticommuting,
ensures that the modified theory is equivalent to 
the original Yang--Mills model. Moreover, the action in the 
exponent (1.8) turns out to be invariant under the
shifted gauge transformations \cite{slav08}
\begin{equation}
\delta A_{\mu}^{a}=\partial_{\mu}\eta^{a}
-g \varepsilon^{abc}A_{\mu}^{b}\eta^{c},
\label{(1.9)}
\end{equation}
\begin{equation}
\delta \varphi^{0}={g\over 2}\varphi^{a}\eta^{a},
\label{(1.10)}
\end{equation}
\begin{equation}
\delta \varphi^{a}=-{a \eta^{a}\over 2}-{g\over 2}
\varepsilon^{abc}\varphi^{b}\eta^{c}
-{g\over 2}\varphi^{0}\eta^{a},
\label{(1.11)}
\end{equation}
\begin{equation}
\delta \chi^{a}={a \eta^{a}\over 2}-{g\over 2}
\varepsilon^{abc}\chi^{b}\eta^{c}
-{g\over 2}\chi^{0}\eta^{a},
\label{(1.12)}
\end{equation}
\begin{equation}
\delta \chi^{0}={g\over 2}\chi^{a}\eta^{a},
\label{(1.13)}
\end{equation}
\begin{equation}
\delta b^{a}=-{g\over 2}\varepsilon^{adc}b^{d}\eta^{c}
-{g\over 2}b^{0}\eta^{a},
\label{(1.14)}
\end{equation}
\begin{equation}
\delta b^{0}={g\over 2}b^{a}\eta^{a},
\label{(1.15)}
\end{equation}
\begin{equation}
\delta c^{a}=-{g\over 2}\varepsilon^{adc}e^{d}\eta^{c},
\label{(1.16)}
\end{equation}
\begin{equation}
\delta e^{0}={g\over 2}e^{a}\eta^{a},
\label{(1.17)}
\end{equation}
where the scalar field $\varphi$ has been represented 
in terms of Hermitian components in the form \cite{slav08}
\begin{equation}
\varphi=\left({i \varphi_{1}+\varphi_{2} \over \sqrt{2}},
{\varphi_{0}-i \varphi_{3}\over \sqrt{2}}\right),
\label{(1.18)}
\end{equation}
and the same for $\chi$.

Section II proves equivalence of two functional-integral formulae
for the S-matrix in the Coulomb gauge. Section III studies the
behaviour of the S-matrix under shift of integration variables.
Section IV contains a detailed proof of invariance of the 
S-matrix under the shifted gauge transformations (1.9)--(1.17).
Concluding remarks and open problems are presented in Sec. V.

\section{Equivalent functional-integral formulae for the S-matrix
in the Coulomb gauge}
 
We start from Eq. (1.4) and recall that,
if $\varphi $ and $\psi $ are commuting and anti-commuting (Grassmann)
complex scalar fields respectively, one has the following functional integral
results:
\begin{equation}
\int d[\varphi ^{\ast }]d[\varphi ]e^{-\int \varphi ^{\ast }B\varphi dx}
=(\det B)^{-1},  
\label{(2.1)}
\end{equation}
\begin{equation}
\int d[\psi ^{\ast }]d[\psi ]e^{-\int \psi ^{\ast }B\psi dx} =(\det B),
\label{(2.2)}
\end{equation}
where $dx\equiv d^{4}x$, and Wick rotation is performed when use is 
made of (2.1) and (2.2).

Equation (1.4) can be therefore expressed as ($\varphi$ and $\varphi^{*}$
being independent in the Euclidean regime)
\begin{equation}
S =S_{0}\times \int d[\varphi^{*}] 
d[\varphi]e^{\int (D_{\mu}\varphi^{*})
(D^{\mu}\varphi)dx}\times ... ,
\label{(2.3)}
\end{equation}
where
\begin{equation}
S_{0} \equiv \int d[\mu ]e^{i\int L_{YM}dx}\delta (\partial_{j}A_{j}).
\label{(2.4)}
\end{equation}
Note now that
\begin{equation}
D_{\mu }\varphi ^{\ast }D^{\mu }\varphi =D_{\mu }(\varphi ^{\ast }D^{\mu
}\varphi )-\varphi ^{\ast }D_{\mu }D^{\mu }\varphi,
\label{(2.5)}
\end{equation}
and the first term, being a total derivative, gives vanishing contribution  
by using the Stokes theorem and  
imposing suitable boundary conditions. Then 
\begin{eqnarray}
S &=&S_{0}\times \int d[\varphi^{\ast }]d[\varphi]
e^{\int (D_{\mu}\varphi^{*})(D^{\mu}\varphi )dx}
\times \int d[\chi ^{\ast }]d[\chi ]e^{-\int
(D_{\mu }\chi^{*})(D^{\mu}\chi )dx}  \nonumber \\
&&\times \int d[b^{\ast }]d[e]
e^{\int (D_{\mu}b^{*})(D^{\mu}e)dx}\times 
\int d[b]d[e]e^{\int (D_{\mu}e^{*})(D^{\mu }b)dx}  
\nonumber \\
&=&S_{0}\times \int d[\varphi^{*}] d[\varphi]
e^{-\int \varphi ^{\ast }D_{\mu }D^{\mu
}\varphi dx}\times \int d[\chi ^{\ast }]d[\chi ]e^{\int \chi ^{\ast }D_{\mu
}D^{\mu }\chi dx}  \nonumber \\
&&\times \int d[b^{\ast }]d[e]e^{-\int b^{\ast }D_{\mu }D^{\mu }edx}\times
\int d[e^{*}]d[b]e^{-\int e^{\ast }D_{\mu }D^{\mu }bdx}  \nonumber \\
&=&S_{0}\times \left\vert \det D^{2}\right\vert ^{-1}\times \left\vert \det
D^{2}\right\vert ^{-1}\times \left\vert \det D^{2}\right\vert \times
\left\vert \det D^{2}\right\vert  \nonumber \\
&=&S_{0}.
\label{(2.6)}
\end{eqnarray}
Note that the independence of the fields $\chi$ and $\chi^{*}$ can be
exploited to consider the rotation $\chi \rightarrow i \chi$,
$\chi^{*} \rightarrow i \chi^{*}$.

\section{Behaviour of the S-matrix under shift of 
integration variables}

In Eq. (1.4) we now perform the shift of integration variables 
described by (1.5)--(1.7), and then point out that
\begin{eqnarray}
(D_{\mu}\varphi^{*})(D^{\mu}\varphi) &\rightarrow &\left( \left(
D_{\mu }\varphi^{*}\right)+g^{-1}
\left( D_{\mu }\hat{a}^{*}\right)\right) 
\left( D^{\mu }\varphi +g^{-1}D^{\mu }\hat{a}\right)  
\nonumber \\
&=&\left( D_{\mu}\varphi^{*} \right)D^{\mu }\varphi +g^{-1}\left(
D_{\mu }\hat{a}^{*}\right)D^{\mu }\varphi 
+g^{-1}\left(D_{\mu }\varphi^{*}\right)(D^{\mu}\hat{a}) 
\nonumber \\
&&+g^{-2}\left( D_{\mu }\hat{a}^{*}\right)D^{\mu}\hat{a}.
\label{(3.1)}
\end{eqnarray}
Thus, if $g^{-2}\rightarrow 0$, one finds
\begin{equation}
(D_{\mu }\varphi^{*})(D^{\mu }\varphi) 
\rightarrow (D_{\mu }\varphi^{*})
(D^{\mu}\varphi)+g^{-1}\left(D_{\mu}\hat{a}^{*}\right) 
(D^{\mu}\varphi)+g^{-1}(D_{\mu}\varphi^{*})(D^{\mu}\hat{a}),
\label{(3.2)}
\end{equation}
\begin{equation}
(D_{\mu}\chi^{*})(D^{\mu }\chi) \rightarrow (D_{\mu }\chi^{*})
(D^{\mu}\chi)-g^{-1}\left(D_{\mu}\hat{a}^{*}\right)(D^{\mu}\chi)
-g^{-1}(D_{\mu }\chi^{*})(D^{\mu }\hat{a}),
\label{(3.3)}
\end{equation}
and the new action defined in Eq. (1.4) transforms as
\begin{eqnarray}
S &\rightarrow &\tilde{S}=\int \exp \{i\int [L_{YM}
+(D_{\mu }\varphi^{*})
(D^{\mu }\varphi)+g^{-1}(D_{\mu }\hat{a}^{*})(D^{\mu}\varphi)
+g^{-1}(D_{\mu}\varphi^{*})(D^{\mu}\hat{a})  
\nonumber \\
&&\ \ \ \ \ \ \ \ \ \ \ \ \ \ \ \ \ \ \ \ \ \ \ 
-(D_{\mu }\chi^{*})(D^{\mu}\chi)
+g^{-1}(D_{\mu}\hat{a}^{*})(D^{\mu}\chi)
+g^{-1}(D_{\mu}\chi^{*})(D^{\mu}\hat{a})  
\nonumber \\
&&\ \ \ \ \ \ \ \ \ \ \ \ \ \ \ \ \ \ \ \ \ \ \ 
+(D_{\mu}b^{*})(D^{\mu}e)
+(D_{\mu}e^{*})(D^{\mu}b)]dx\ \}\delta (\partial_{j}A_{j})
d\gamma.
\label{(3.4)}
\end{eqnarray}

\section{Invariance of the S-matrix under shifted gauge transformations}

We are aiming to show that 
the transformed action in Eq. (3.4) is
invariant under the shifted gauge transformations (1.9)--(1.17).
To begin, note that the covariant derivative in the
spinor representation reads as 
\begin{equation}
D^{\mu}\varphi =\partial ^{\mu}\varphi -ig\sigma
^{a}A^{a\mu}\varphi,
\label{(4.1)}
\end{equation}
\begin{equation}
D_{\mu }\varphi^{*} =\partial _{\mu }\varphi ^{\ast }+ig\varphi
^{\ast }\sigma ^{a}A_{\mu }^{a}, 
\label{(4.2)}
\end{equation}
\begin{eqnarray}
(D_{\mu}\varphi^{*})\left( D^{\mu }\varphi 
\right) &=&\left( \partial
_{\mu }\varphi ^{\ast }+ig\varphi ^{\ast }\sigma ^{a}A_{\mu }^{a}\right)
\left( \partial ^{\mu }\varphi -ig\sigma ^{a}A^{a\mu }
\varphi \right)  \nonumber \\
&=&\partial _{\mu }\varphi ^{\ast }\partial ^{\mu }\varphi +ig\left( \varphi
^{\ast }\sigma ^{a}\partial ^{\mu }\varphi \right) A_{\mu }^{a}-ig\left(
\partial _{\mu }\varphi ^{\ast }\sigma ^{a}\varphi \right) A^{a\mu }  
\nonumber \\
&&+g^{2}\left( \varphi ^{\ast }\sigma ^{a}\sigma ^{b}\varphi \right) A_{\mu
}^{a}A^{b\mu}.
\label{(4.3)}
\end{eqnarray}
The Pauli matrices, $\varphi $ and $\varphi ^{\ast }$ 
are all Hermitian in the Minkowskian regime, therefore
\begin{equation}
\varphi ^{\ast }\sigma ^{a}\partial ^{\mu }\varphi =\left( \varphi ^{\ast
}\sigma ^{a}\partial ^{\mu }\varphi \right) ^{\dagger }=\partial ^{\mu
}\varphi ^{\ast }\sigma ^{a}\varphi .
\label{(4.4)}
\end{equation}
In the Euclidean regime, however, which is necessary to obtain
well defined functional integrals, $\varphi$ and $\varphi^{*}$
become independent (see also comment before (2.3)), not related
by any conjugation, despite being denoted in the same way.
Moreover, we exploit the identities
\begin{equation}
\varphi ^{\ast }\sigma ^{a}\sigma ^{b}\varphi 
=\varphi ^{\ast }\mathbf{I}
\delta ^{ab}\varphi +i\varepsilon ^{abc}\varphi ^{\ast }
\sigma ^{c}\varphi, 
\label{(4.5)}
\end{equation}
\begin{equation}
\varepsilon ^{abc}A_{\mu }^{a}A^{b\mu } =0,
\label{(4.6)}
\end{equation}
and hence
\begin{equation}
(D_{\mu}\varphi^{*})\left( D^{\mu }\varphi 
\right) =\left( \partial
_{\mu }\varphi ^{\ast }\right) \left( \partial ^{\mu }\varphi \right)
+g^{2}\varphi ^{\ast }\varphi A_{\mu }^{a}A^{a\mu }.
\label{(4.7)}
\end{equation}

We notice also that 
\begin{equation}
\partial _{\mu }\varphi ^{\ast }\partial ^{\mu }\varphi =\left( \partial
_{\mu }\varphi ^{a}\right) \left( \partial ^{\mu }\varphi ^{a}\right)
+\left( \partial _{\mu }\varphi ^{0}\right) \left( \partial ^{\mu }\varphi
^{0}\right),
\label{(4.8)}
\end{equation}
\begin{equation}
\varphi ^{\ast }\varphi =\varphi ^{a}\varphi ^{a}+\varphi ^{0}\varphi^{0},
\label{(4.9)}
\end{equation}
which implies
\begin{equation}
(D_{\mu}\varphi^{*})\left( D^{\mu }\varphi \right) =\partial
_{\mu }\varphi ^{a}\partial ^{\mu }\varphi ^{a}+\partial _{\mu }\varphi
^{0}\partial ^{\mu }\varphi ^{0}+g^{2}\left( \varphi ^{a}\varphi
^{a}+\varphi ^{0}\varphi ^{0}\right) A_{\mu }^{a}A^{a\mu },
\label{(4.10)}
\end{equation}
and similarly 
\begin{equation}
(D_{\mu }\chi^{*})\left( D^{\mu }\chi \right) 
=\partial _{\mu
}\chi ^{a}\partial ^{\mu }\chi ^{a}+\partial _{\mu }\chi ^{0}\partial ^{\mu
}\chi ^{0}+g^{2}\left( \chi ^{a}\chi ^{a}+\chi ^{0}\chi ^{0}\right) A_{\mu
}^{a}A^{a\mu }.
\label{(4.11)}
\end{equation}

Let us now explicitly express the other terms in the 
action Eq. (3.4), i.e.
\begin{eqnarray}
g^{-1}(D_{\mu }\hat{a})^{\ast }(D^{\mu }\varphi ) 
&=&g^{-1}(\partial _{\mu }
\hat{a}^{\ast }+ig\hat{a}^{\ast }\sigma ^{a}A_{\mu }^{a})(\partial ^{\mu
}\varphi -ig\sigma ^{a}A^{a\mu }\varphi )  \nonumber \\
&=&i\hat{a}^{\ast }\sigma ^{a}A_{\mu }^{a}(\partial ^{\mu }
\varphi -ig\sigma
^{a}A^{a\mu }\varphi )  \nonumber \\
&=&i\left( \hat{a}^{\ast }\sigma ^{a}\partial ^{\mu }
\varphi \right) A_{\mu
}^{a}+g\left( \hat{a}^{\ast }\sigma ^{a}\sigma ^{b}\varphi \right) A_{\mu
}^{a}A^{b\mu }  \nonumber \\
&=&i\left( \hat{a}^{\ast }\sigma^{a}\partial ^{\mu }\varphi \right) A_{\mu
}^{a}+g\left( \hat{a}^{\ast }\varphi \right) A_{\mu }^{a}A^{a\mu },
\label{(4.12)}
\end{eqnarray}
\begin{eqnarray}
g^{-1}(D_{\mu }\varphi^{*})(D^{\mu }\hat{a}) &=&g^{-1}\left( \partial
_{\mu }\varphi ^{\ast }+ig\varphi ^{\ast }\sigma ^{a}A_{\mu }^{a}\right)
(\partial ^{\mu }\hat{a}-ig\sigma ^{a}A^{a\mu }\hat{a})  \nonumber \\
&=&-i\left( \partial _{\mu }\varphi ^{\ast }\sigma ^{a}\hat{a}\right)
A^{a\mu }+g\left( \varphi ^{\ast }\sigma ^{a}\sigma ^{b}\hat{a}\right)
A_{\mu }^{a}A^{b\mu }  \nonumber \\
&=&-i\left( \partial _{\mu }\varphi ^{\ast }\sigma ^{a}\hat{a}\right)
A^{a\mu }+g\left( \varphi ^{\ast }\hat{a}\right) A_{\mu }^{a}A^{a\mu}.
\label{(4.13)}
\end{eqnarray}
At this stage, the Hermiticity condition
\begin{equation}
\left( \hat{a}^{\ast }\sigma ^{a}\partial ^{\mu }\varphi \right) =\left( 
\hat{a}^{\ast }\sigma ^{a}\partial ^{\mu }\varphi \right) ^{\dagger }=\left(
\partial _{\mu }\varphi ^{\ast }\sigma ^{a}\hat{a}\right)
\label{(4.14)}
\end{equation}
and the previous formulae lead to
\begin{eqnarray*}
g^{-1}(D_{\mu }\hat{a})^{\ast }(D^{\mu }\varphi )
+g^{-1}(D_{\mu}\varphi^{*})(D^{\mu}
\hat{a}) &=&g\left( \hat{a}^{\ast }\varphi \right) A_{\mu
}^{a}A^{a\mu }+g\left( \varphi ^{\ast }\hat{a}\right) A_{\mu }^{a}A^{a\mu }
\\
&=&gA_{\mu }^{a}A^{a\mu }\left( \hat{a}^{\ast }\varphi 
+\varphi ^{\ast }\hat{
a}\right) =ag\varphi ^{0}A_{\mu }^{a}A^{a\mu },
\end{eqnarray*}
which implies
\begin{equation}
g^{-1}(D_{\mu }\hat{a})^{\ast }(D^{\mu }\varphi )
+g^{-1}(D_{\mu}\varphi^{*})(D^{\mu}\hat{a})
=ag\varphi^{0}A_{\mu }^{a}A^{a\mu}.
\label{(4.15)}
\end{equation}

In the same way one gets 
\begin{equation}
(D_{\mu }b^{*})(D^{\mu}e)+(D_{\mu}e^{*})(D^{\mu}b)
=\partial _{\mu
}b^{\ast }\partial ^{\mu }e+\partial _{\mu }e^{\ast }\partial ^{\mu
}b+g^{2}\left( b^{\ast }e+e^{\ast }b\right) A_{\mu }^{a}A^{a\mu},
\label{(4.16)}
\end{equation}
or, equivalently,
\begin{equation}
(D_{\mu}b^{*})(D^{\mu}e)+(D_{\mu}e^{*})(D^{\mu}b)
=\partial _{\mu }b^{0}\partial ^{\mu}e^{0}+\partial _{\mu }e^{i}\partial
^{\mu }b^{i}+g^{2}\left( b^{0}e^{0}+b^{i}e^{i}\right) 
A_{\mu }^{a}A^{a\mu}.
\label{(4.17)}
\end{equation}

By virtue of (4.10), (4.11), (4.15) and (4.17) the action in (3.4)
takes the form
\begin{eqnarray}
L_{{\rm tot}} &=&L_{YM}+(D_{\mu }\varphi^{*})
(D^{\mu }\varphi)+g^{-1}(D_{\mu}\hat{a}^{*})
(D^{\mu }\varphi)
+g^{-1}(D_{\mu }\varphi^{*})(D^{\mu}\hat{a})  
\nonumber \\
&&\ \ \ \ \ \ \ \ -(D_{\mu }\chi^{*})(D^{\mu }\chi )+g^{-1}(D_{\mu }
\hat{a}^{*})(D^{\mu}\chi)
+g^{-1}(D_{\mu}\chi^{*})(D^{\mu}\hat{a})  
\nonumber \\
&&\ \ \ \ \ \ \ \ +(D_{\mu }b^{*})(D^{\mu}e)
+(D_{\mu}e^{*})(D^{\mu}b)  
\nonumber \\
&&=-\frac{1}{4}F_{\mu \nu }^{a}F^{a\mu \nu }  \nonumber \\
&&+\left( \partial _{\mu }\varphi ^{a}\partial ^{\mu }\varphi ^{a}+\partial
_{\mu }\varphi ^{0}\partial ^{\mu }\varphi ^{0}\right) +g^{2}\left( \varphi
^{a}\varphi ^{a}+\varphi ^{0}\varphi ^{0}\right) A_{\mu}^{a}A^{a\mu} 
\nonumber \\
&&+ag\varphi ^{0}A_{\mu }^{a}A^{a\mu }  \nonumber \\
&&-\left( \partial _{\mu }\chi ^{a}\partial ^{\mu }\chi ^{a}
+\partial _{\mu
}\chi ^{0}\partial ^{\mu }\chi ^{0}\right) -g^{2}\left( \chi ^{a}\chi
^{a}+\chi ^{0}\chi ^{0}\right) A_{\mu }^{a}A^{a\mu }  \nonumber \\
&&+ag\chi ^{0}A_{\mu }^{a}A^{a\mu }  \nonumber \\
&&+\left( \partial _{\mu }b^{0}\partial ^{\mu }e^{0}+\partial ^{\mu
}b^{a}\partial _{\mu }e^{a}\right) +g^{2}\left( b^{0}e^{0}
+b^{a}e^{a}\right)
A_{\mu }^{c}A^{c\mu}.
\label{(4.18)}
\end{eqnarray}

Now we just list the equations of motion for all fields 
in the action pertaining to the Lagrangian (4.18), i.e.
$
\partial_{\mu}\frac{\partial L_{{\rm tot}}}
{\partial \left( \partial_{\mu }A_{\nu}^{a}\right) }
-\frac{\partial L_{{\rm tot}}}{\partial 
\left( A_{\nu}^{a}\right) }
=0\rightarrow
$
\begin{equation}
D_{\mu }F^{c\mu \nu }=-2g^{2}A^{c\nu }\left( \varphi ^{a}\varphi
^{a}+\chi ^{a}\chi ^{a}+b^{a}e^{a}+\varphi ^{0}\varphi ^{0}+\chi ^{0}\chi
^{0}+b^{0}e^{0}\right) -2agA^{c\nu }\left(\varphi^{0}+\chi^{0}\right),
\label{(4.19)}
\end{equation}
$$
\partial _{\mu }\frac{\partial L_{{\rm tot}}}
{\partial \left( \partial _{\mu }\varphi
^{a}\right) }-\frac{\partial L_{{\rm tot}}}{\partial 
\left( \varphi ^{a}\right)}
=0\rightarrow
$$
\begin{equation}
\partial ^{2}\varphi ^{a}=g^{2}\varphi ^{a}A_{\mu }^{b}A^{b\mu},
\label{(4.20)}
\end{equation}
\begin{equation}
\partial ^{2}\varphi ^{0}=\left( g^{2}\varphi ^{0}+\frac{ag}{2}
\right) A_{\mu}^{a}A^{a\mu},
\label{(4.21)}
\end{equation}
\begin{equation}
\partial ^{2}\chi ^{a}=g^{2}\chi ^{a}A_{\mu }^{b}A^{b\mu},
\label{(4.22)}
\end{equation}
\begin{equation}
\partial ^{2}\chi ^{0}=\left( g^{2}\chi ^{0}-\frac{ag}{2}\right)
A_{\mu }^{a}A^{a\mu},
\label{(4.23)}
\end{equation}
\begin{equation}
\partial^{2}e^{a}=g^{2}e^{a}A_{\mu }^{b}A^{b\mu},
\label{(4.24)}
\end{equation}
\begin{equation}
\partial^{2}e^{0}=g^{2}e^{0}A_{\mu}^{b}A^{b\mu},
\label{(4.25)}
\end{equation}
\begin{equation}
\partial ^{2}b^{a}=g^{2}b^{a}A_{\mu }^{c}A^{c\mu },
\label{(4.26)}
\end{equation}
\begin{equation}
\partial ^{2}b^{0}=g^{2}b^{0}A_{\mu }^{c}A^{c\mu}.
\label{(4.27)}
\end{equation}
The variation of the action obtained from Eq. (4.18) is
\begin{eqnarray}
\delta L_{{\rm tot}} &=&\delta \left( 
-\frac{1}{4}F_{\mu \nu }^{a}F^{a\mu \nu
}\right)  \nonumber \\
&&+2\left( \partial _{\mu }\delta \varphi ^{a}\partial ^{\mu }\varphi
^{a}+\partial _{\mu }\delta \varphi ^{0}\partial ^{\mu }\varphi ^{0}\right)
+2g^{2}\left( \delta \varphi ^{a}\varphi ^{a}+\delta \varphi ^{0}\varphi
^{0}\right) A_{\mu }^{a}A^{a\mu }  \nonumber \\
&&\ \ \ \ \ \ \ \ \ \ \ \ \ \ \ \ \ \ \ \ \ \ \ \ \ \ \ \ +2g^{2}\left(
\varphi ^{a}\varphi ^{a}+\varphi ^{0}\varphi ^{0}\right) \delta A_{\mu
}^{a}A^{a\mu }  \nonumber \\
&&+ag\delta \varphi ^{0}A_{\mu }^{a}A^{a\mu }+2ag\varphi ^{0}
\delta A_{\mu
}^{a}A^{a\mu }  \nonumber \\
&&-2\left( \partial _{\mu }\delta \chi ^{a}\partial ^{\mu }\chi
^{a}+\partial _{\mu }\delta \chi ^{0}\partial ^{\mu }\chi ^{0}\right)
-2g^{2}\left( \delta \chi ^{a}\chi ^{a}+\delta \chi ^{0}\chi ^{0}\right)
A_{\mu }^{a}A^{a\mu }  \nonumber \\
&&\ \ \ \ \ \ \ \ \ \ \ \ \ \ \ \ \ \ \ \ \ \ \ \ \ \ \ \ -2g^{2}\left( \chi
^{a}\chi ^{a}+\chi ^{0}\chi ^{0}\right) \delta A_{\mu }^{a}A^{a\mu }  
\nonumber \\
&&+ag\delta \chi ^{0}A_{\mu }^{a}A^{a\mu }+2ag\chi ^{0}\delta A_{\mu
}^{a}A^{a\mu }  \nonumber \\
&&+\left( \partial _{\mu }\delta b^{0}\partial ^{\mu }e^{0}+\partial _{\mu
}b^{0}\partial ^{\mu }\delta e^{0}+\partial ^{\mu }b^{d}\partial _{\mu
}\delta e^{d}+\partial ^{\mu }\delta b^{d}\partial _{\mu }e^{d}\right) 
\nonumber \\
&&\ \ \ \ \ \ \ \ \ \ \ \ \ \ \ \ \ \ \ \ \ \ \ \ \ \ \ \ +g^{2}\left(
\delta b^{0}e^{0}+b^{0}\delta e^{0}+\delta b^{i}e^{i}+b^{i}\delta
e^{i}\right) A_{\mu }^{a}A^{a\mu }  \nonumber \\
&&\ \ \ \ \ \ \ \ \ \ \ \ \ \ \ \ \ \ \ \ \ \ \ \ \ \ \ \ +2g^{2}\left(
b^{0}e^{0}+b^{i}e^{i}\right) \delta A_{\mu }^{a}A^{a\mu}.
\label{(4.28)}
\end{eqnarray}

Variation of the gauge-field Lagrangian $L_{YM}$, 
separately, is as follows:
\begin{equation}
\delta L_{YM} =-\frac{1}{4}\delta (F_{\mu \nu }^{a}
F^{a\mu \nu })=-\frac{1
}{2}F^{a\mu \nu }\delta F_{\mu \nu }^{a},
\label{(4.29)}
\end{equation}
\begin{equation}
\delta F_{\mu \nu }^{a} =\partial _{\mu }\delta A_{\nu }^{a}-\partial
_{\nu }\delta A_{\mu }^{a}+g\varepsilon ^{abc}\delta A_{\mu }^{b}A_{\nu
}^{c}+g\varepsilon ^{abc}A_{\mu }^{b}\delta A_{\nu }^{c},
\label{(4.30)}
\end{equation}
\begin{eqnarray}
\Longrightarrow F^{a\mu \nu }\delta F_{\mu \nu }^{a}&=&F^{a\mu \nu
}(\partial _{\mu }\delta A_{\nu }^{a}-\partial _{\nu }\delta A_{\mu
}^{a}+g\varepsilon ^{abc}\delta A_{\mu }^{b}A_{\nu }^{c}+g\varepsilon
^{abc}A_{\mu }^{b}\delta A_{\nu }^{c})  \nonumber \\
&=&2F^{a\mu \nu }(\partial _{\mu }\delta A_{\nu }^{a}+g\varepsilon
^{abc}A_{\mu }^{b}\delta A_{\nu }^{c}),
\label{(4.31)}
\end{eqnarray}
\begin{equation}
\delta A_{\nu }^{a} =\partial _{\nu }\eta ^{a}-g\varepsilon ^{abc}A_{\nu
}^{b}\eta ^{c} ,
\label{(4.32)}
\end{equation}
\begin{eqnarray}
\delta L_{YM} &=&-F^{a\mu \nu }\left( \partial _{\mu }
\left( \partial _{\nu
}\eta ^{a}-g\varepsilon ^{abc}A_{\nu }^{b}\eta ^{c}\right) +g\varepsilon
^{abc}A_{\mu }^{b}\left( \partial _{\nu }\eta ^{c}
-g\varepsilon ^{cde}A_{\nu}^{d}\eta^{e}\right) \right) \nonumber \\
&=&-F^{a\mu \nu }\Bigr( \partial_{\mu }\partial_{\nu }\eta
^{a}-g\varepsilon ^{abc}\partial _{\mu }\left( A_{\nu }^{b}
\eta ^{c}\right) \nonumber \\
&+& g\varepsilon ^{abc}A_{\mu }^{b}\partial _{\nu }
\eta ^{c}-g^{2}\varepsilon^{abc}
\varepsilon ^{cde}A_{\mu }^{b}A_{\nu }^{d}\eta ^{e}\Bigr).
\label{(4.33)}
\end{eqnarray}
In light of the identities
\begin{equation}
F^{a\mu \nu }\partial _{\mu }\partial _{\nu }\eta ^{a} =0,
\label{(4.34)}
\end{equation}
\begin{equation}
\varepsilon ^{abc}\varepsilon ^{cde} =\varepsilon ^{cab}\varepsilon
^{cde}=\delta ^{ae}\delta ^{bd}-\delta ^{ad}\delta ^{be},
\label{(4.35)}
\end{equation}
one finds eventually
\begin{equation}
\delta L_{YM}=g\varepsilon ^{abc}F^{a\mu \nu }\partial _{\mu }\left(
A_{\nu }^{b}\eta ^{c}\right) -g\varepsilon ^{abc}F^{a\mu \nu }A_{\mu
}^{b}\partial _{\nu }\eta ^{c}+g^{2}F^{a\mu \nu}
A_{\mu }^{b}A_{\nu}^{a}\eta^{b},
\label{(4.36)}
\end{equation}
Thus, Eqs. (1.9)--(1.17) and (3.4) yield
\begin{eqnarray}
\delta L_{{\rm tot}} &=&g\varepsilon^{abc}
F^{a\mu \nu }\partial _{\mu }\left(
A_{\nu }^{b}\eta ^{c}\right) -g\varepsilon ^{abc}F^{a\mu \nu }A_{\mu
}^{b}\partial _{\nu }\eta ^{c}+g^{2}F^{a\mu \nu }A_{\mu }^{b}A_{\nu
}^{a}\eta ^{b}  \nonumber \\
&&+2\partial _{\mu }\left( -\frac{a\eta ^{a}}{2}-\frac{g}{2}\varepsilon
^{abc}\varphi ^{b}\eta ^{c}-\frac{g}{2}\varphi ^{0}
\eta ^{a}\right) \partial
^{\mu }\varphi ^{a}+2\partial _{\mu }\left( \frac{g}{2}\varphi ^{a}\eta
^{a}\right) \partial ^{\mu }\varphi ^{0}  \nonumber \\
&&+2g^{2}A_{\mu }^{d}A^{d\mu }\left( -\frac{a\eta ^{a}}{2}-\frac{g}{2}
\varepsilon ^{abc}\varphi ^{b}\eta ^{c}-\frac{g}{2}\varphi ^{0}\eta
^{a}\right) \varphi ^{a}+2g^{2}A_{\mu }^{b}A^{b\mu }\left( \frac{g}{2}
\varphi ^{a}\eta ^{a}\right) \varphi ^{0}  \nonumber \\
&&+2g^{2}\left( \varphi ^{d}\varphi ^{d}+\varphi ^{0}\varphi ^{0}\right)
\left( \partial _{\mu }\eta ^{a}-g\varepsilon ^{abc}A_{\mu }^{b}\eta
^{c}\right) A^{a\mu }  \nonumber \\
&&+ag\left( \frac{g}{2}\varphi ^{a}\eta ^{a}\right) A_{\mu }^{b}A^{b\mu
}+2ag\varphi ^{0}\left( \partial _{\mu }\eta ^{a}
-g\varepsilon ^{abc}A_{\mu
}^{b}\eta ^{c}\right) A^{a\mu }  \nonumber \\
&&-2\partial _{\mu }\left( \frac{a\eta ^{a}}{2}-\frac{g}{2}\varepsilon
^{abc}\chi ^{b}\eta ^{c}-\frac{g}{2}\chi ^{0}\eta ^{a}\right) 
\partial ^{\mu
}\chi ^{a}-2\partial _{\mu }\left( \frac{g}{2}\chi ^{a}\eta ^{a}\right)
\partial ^{\mu }\chi ^{0}  \nonumber \\
&&-2g^{2}\left( \frac{a\eta ^{a}}{2}-\frac{g}{2}\varepsilon ^{abc}\chi
^{b}\eta ^{c}-\frac{g}{2}\chi ^{0}\eta ^{a}\right) \chi ^{a}A_{\mu
}^{d}A^{d\mu }-2g^{2}\left( \frac{g}{2}\chi ^{a}\eta ^{a}\right) \chi
^{0}A_{\mu }^{b}A^{b\mu }  \nonumber \\
&&+2g^{2}\left( \chi ^{d}\chi ^{d}+\chi ^{0}\chi ^{0}
\right) \left( \partial
_{\mu }\eta ^{a}-g\varepsilon ^{abc}A_{\mu }^{b}
\eta ^{c}\right) A^{a\mu } 
\nonumber \\
&&+ag\left( \frac{g}{2}\chi ^{a}\eta ^{a}\right) A_{\mu }^{b}A^{b\mu
}+2ag\chi ^{0}\left( \partial _{\mu }\eta ^{a}-g\varepsilon ^{abc}A_{\mu
}^{b}\eta ^{c}\right) A^{a\mu }  \nonumber \\
&&+\partial _{\mu }\left( \frac{g}{2}b^{a}\eta ^{a}\right) \partial ^{\mu
}e^{0}+\partial _{\mu }b^{0}\partial ^{\mu }\left( \frac{g}{2}e^{a}\eta
^{a}\right) +\partial ^{\mu }b^{a}\partial _{\mu }\left( -\frac{g}{2}
\varepsilon ^{adc}e^{d}\eta ^{c}\right)  \nonumber \\
&&+\partial ^{\mu }\left( -\frac{g}{2}\varepsilon ^{adc}b^{d}
\eta ^{c}-\frac{
g}{2}b^{0}\eta ^{a}\right) \partial _{\mu }e^{a}  \nonumber \\
&&+g^{2}A_{\mu }^{b}A^{b\mu }\left( \frac{g}{2}b^{a}\eta ^{a}\right)
e^{0}+g^{2}A_{\mu }^{b}A^{b\mu }b^{0}\left( \frac{g}{2}e^{a}\eta ^{a}\right)
\nonumber \\
&&+g^{2}A_{\mu }^{b}A^{b\mu }\left( -\frac{g}{2}\varepsilon ^{adc}b^{d}\eta
^{c}-\frac{g}{2}b^{0}\eta ^{a}\right) e^{a}+g^{2}A_{\mu }^{b}A^{b\mu
}b^{a}\left( -\frac{g}{2}\varepsilon ^{adc}
e^{d}\eta^{c}\right)  \nonumber \\
&&+2g^{2}\left( b^{0}e^{0}+b^{d}e^{d}\right) \left( \partial _{\mu }\eta
^{a}-g\varepsilon ^{abc}A_{\mu }^{b}\eta^{c}\right) A^{a\mu}.
\label{(4.37)}
\end{eqnarray}
This is a huge Lagrangian variation that we should show 
is identically vanishing! For this purpose, we write in detail
all terms in the form
\begin{eqnarray}
\delta L_{{\rm tot}} &=&g\varepsilon^{abc}
F^{a\mu \nu }\partial _{\mu }\left(
A_{\nu }^{b}\eta ^{c}\right) -g\varepsilon ^{abc}F^{a\mu \nu }A_{\mu
}^{b}\partial _{\nu }\eta ^{c}+g^{2}F^{a\mu \nu }A_{\mu }^{b}A_{\nu
}^{a}\eta ^{b}  \nonumber \\
&&-{\underbrace{a\partial _{\mu }\eta ^{a}\partial ^{\mu
}\varphi ^{a}}}-{\underbrace{g\varepsilon ^{abc}\partial _{\mu
}\left( \varphi ^{b}\eta ^{c}\right) \partial ^{\mu }\varphi ^{a}}}
-{\underbrace{g\partial _{\mu }\left( \varphi ^{0}\eta ^{a}\right)
\partial ^{\mu }\varphi ^{a}}}+{\underbrace{g\partial _{\mu
}\left( \varphi ^{a}\eta ^{a}\right) \partial ^{\mu }\varphi ^{0}}}  
\nonumber \\
&&-{\underbrace{ag^{2}A_{\mu }^{b}A^{b\mu }
\varphi ^{a}\eta ^{a}}}
-{\underbrace{g^{3}\varepsilon ^{abc}A_{\mu }^{d}A^{d\mu}
\varphi ^{a}\varphi ^{b}\eta ^{c}}}-{\underbrace{g^{3}A_{\mu
}^{b}A^{b\mu }\varphi ^{a}\eta ^{a}\varphi ^{0}}}+{\underbrace{
g^{3}A_{\mu }^{b}A^{b\mu }\varphi ^{a}\eta ^{a}\varphi ^{0}}}  
\nonumber \\
&&+2g^{2}\left( \varphi ^{d}\varphi ^{d}+\varphi ^{0}\varphi ^{0}\right)
\left( \partial _{\mu }\eta ^{a}-g\varepsilon ^{abc}A_{\mu }^{b}\eta
^{c}\right) A^{a\mu }  \nonumber \\
&&+{\underbrace{\frac{ag^{2}}{2}A_{\mu }^{b}A^{b\mu }\varphi
^{a}\eta ^{a}}}+2agA^{a\mu }\partial _{\mu }\eta ^{a}\varphi^{0}
-{\underbrace{2ag^{2}\varepsilon ^{abc}A^{a\mu }A_{\mu }^{b}\eta
^{c}\varphi ^{0}}}  \nonumber \\
&&-{\underbrace{a\partial_{\mu }
\eta ^{a}\partial ^{\mu }\chi
^{a}}}+{\underbrace{g\varepsilon ^{abc}
\partial _{\mu }\left(
\chi ^{b}\eta ^{c}\right) \partial ^{\mu }\chi ^{a}}}
+{\underbrace{g\partial _{\mu }\left( \chi ^{0}\eta ^{a}
\right) \partial ^{\mu
}\chi ^{a}}}-{\underbrace{g\partial_{\mu}
\left( \chi ^{a}\eta
^{a}\right) \partial^{\mu }\chi^{0}}}  \nonumber \\
&&-{\underbrace{ag^{2}\chi^{a}
\eta^{a}A_{\mu}^{b}A^{b\mu}}}+
{\underbrace{g^{3}\varepsilon^{abc}
\chi^{a}\chi^{b}\eta^{c}
A_{\mu}^{d}A^{d\mu}}}  + 
{\underbrace{g^{3}\chi ^{a}\eta
^{a}\chi ^{0}A_{\mu }^{d}A^{d\mu }}}
-{\underbrace{g^{3}\chi
^{a}\eta ^{a}\chi ^{0}A_{\mu}^{b}A^{b\mu}}}  \nonumber \\
&&+2g^{2}\left( \chi ^{d}\chi ^{d}+\chi ^{0}\chi ^{0}
\right) \left( \partial
_{\mu }\eta ^{a}-g\varepsilon ^{abc}A_{\mu }^{b}
\eta ^{c}\right) A^{a\mu } 
\nonumber \\
&&+{\underbrace{\frac{ag^{2}}{2}A_{\mu }^{b}A^{b\mu }\chi
^{a}\eta ^{a}}}+2ag\chi ^{0}A^{a\mu }\partial_{\mu }\eta ^{a}
-{\underbrace{2ag^{2}\varepsilon ^{abc}A^{a\mu }A_{\mu }^{b}
\eta ^{c}\chi ^{0}}}  
\nonumber \\
&&+{\underbrace{\frac{g}{2}\partial_{\mu }\left( b^{a}\eta
^{a}\right) \partial ^{\mu }e^{0}}}+{\underbrace{\frac{g}{2}
\partial_{\mu }b^{0}\partial^{\mu}\left(e^{a}\eta^{a}\right)}}-
{\underbrace{\frac{g}{2}\varepsilon ^{adc}\partial^{\mu
}b^{a}\partial _{\mu }\left( e^{d}\eta ^{c}\right)}}  \nonumber \\
&&-{\underbrace{\frac{g}{2}\varepsilon ^{adc}\partial ^{\mu
}\left( b^{d}\eta ^{c}\right) \partial _{\mu }e^{a}}}
-{\underbrace{\frac{g}{2}\partial^{\mu}\left(b^{0}\eta^{a}
\right) \partial
_{\mu }e^{a}}}  \nonumber \\
&&+{\underbrace{\frac{g^{3}}{2}A_{\mu }^{b}A^{b\mu}
b^{a}e^{0}\eta^{a}}}
+{\underbrace{\frac{g^{3}}{2}A_{\mu
}^{b}A^{b\mu }b^{0}e^{a}\eta ^{a}}}  \nonumber \\
&&-{\underbrace{\frac{g^{3}}{2}\varepsilon^{adc}A_{\mu
}^{b}A^{b\mu }b^{d}e^{a}\eta ^{c}}}
-{\underbrace{\frac{g^{3}}{2}
A_{\mu }^{b}A^{b\mu }b^{0}e^{a}}\eta^{a}}
-{\underbrace{\frac{g^{3}}{2}\varepsilon ^{adc}A_{\mu}^{b}A^{b\mu}b^{a}
e^{d}\eta ^{c}}}  \nonumber \\
&&+2g^{2}\left( b^{0}e^{0}+b^{d}e^{d}\right) \left( \partial _{\mu }\eta
^{a}-g\varepsilon ^{abc}A_{\mu }^{b}\eta ^{c}\right) A^{a\mu}.
\label{(4.38)}
\end{eqnarray}
The $30$ terms underlined by a curly bracket in (4.38) are hereafter
denoted by $T_{i}$, with $i$ ranging from $1$ through $30$. 
To begin, note that, by virtue of (4.20), 
\begin{eqnarray}
T_{1} &=&-a\partial _{\mu}\eta ^{a}\partial ^{\mu}
\varphi^{a}=-
{\underbrace{a\partial _{\mu }\left( \eta ^{a}\partial
^{\mu }\varphi^{a}\right) }}+a\eta ^{a}\partial _{\mu }\partial ^{\mu
}\varphi ^{a}\equiv a\eta ^{a}\partial ^{2}\varphi^{a}  \nonumber \\
&=&ag^{2}\eta ^{a}\varphi^{a}A_{\mu}^{b}A^{b\mu},
\label{(4.39)}
\end{eqnarray}
and hence
\begin{equation}
T_{1}+T_{5}=0.
\label{(4.40)}
\end{equation}
Moreover, again by virtue of (4.20),
\begin{eqnarray}
T_{2} &=&g\varepsilon ^{abc}\partial _{\mu }
\left( \varphi ^{b}\eta ^{c}\right)
\partial ^{\mu }\varphi ^{a}=g\varepsilon ^{abc}\partial _{\mu }\left(
\varphi ^{b}\eta ^{c}\partial ^{\mu }\varphi ^{a}\right) -g\varepsilon
^{abc}\varphi ^{b}\eta ^{c}\partial _{\mu }\partial ^{\mu }\varphi ^{a} 
\nonumber \\
&\equiv &-g\varepsilon^{abc}\varphi^{b}\eta^{c}
\partial^{2}\varphi^{a}=-g^{3}A_{\mu}^{b}A^{b\mu}
{\underbrace{\varepsilon^{abc}\varphi^{b}\varphi^{a}}}\eta^{c},
\label{(4.41)}
\end{eqnarray}
i.e.
\begin{equation}
T_{2}=0,
\label{(4.42)}
\end{equation}
and
\begin{eqnarray}
T_{3} &=&-g\partial _{\mu }\left( \varphi ^{0}\eta ^{a}
\right) \partial ^{\mu
}\varphi ^{a}=-g\partial _{\mu }\left( \varphi ^{0}
\eta ^{a}\partial ^{\mu
}\varphi ^{a}\right) +g\left( \varphi ^{0}\eta ^{a}
\right) \partial _{\mu
}\partial ^{\mu }\varphi ^{a}\equiv g\varphi ^{0}\eta ^{a}\partial
^{2}\varphi ^{a}  \nonumber \\
&&=g^{3}A_{\mu }^{b}A^{b\mu }\varphi^{0}
\eta^{a}\varphi^{a},
\label{(4.43)}
\end{eqnarray}
\begin{equation}
T_{3}+T_{7}=0.
\label{(4.44)}
\end{equation}
The term $T_{4}$ is studied with the help of (4.21), so that
\begin{eqnarray}
T_{4} &=&g\partial _{\mu }\left( \varphi ^{a}\eta ^{a}\right) \partial ^{\mu
}\varphi^{0}=g\partial _{\mu }\left( \varphi ^{a}\eta ^{a}\partial ^{\mu
}\varphi ^{0}\right) -g\varphi ^{a}\eta ^{a}\partial _{\mu }\partial ^{\mu
}\varphi ^{0}\equiv -g\varphi ^{a}\eta ^{a}\partial ^{2}
\varphi ^{0}  \nonumber \\
&&=-g\varphi^{a}\eta^{a}
\left( g^{2}\varphi ^{0}+\frac{ag}{
2}\right) A_{\mu}^{a}A^{a\mu},
\label{(4.45)}
\end{eqnarray}
\begin{equation}
T_{4}+T_{8}+T_{9}=0.
\label{(4.46)}
\end{equation}
Furthermore,
\begin{equation}
T_{6}=-g^{3}{\underbrace{\varepsilon ^{abc}
\varphi^{a}\varphi^{b}}
}A_{\mu }^{d}A^{d\mu }\eta^{c},
\label{(4.47)}
\end{equation}
which vanishes because the antisymmetric $\varepsilon^{abc}$ is
contracted with the symmetric product $\varphi^{a}\varphi^{b}$:
\begin{equation}
T_{6}=0.
\label{(4.48)}
\end{equation}
For the same reason the term $T_{10}$ vanishes as well, 
\begin{equation}
T_{10}=0.
\label{(4.49)}
\end{equation}
Now we consider the terms which include the field $\chi$. By virtue of
(4.22), one finds
\begin{eqnarray}
T_{11} &=&-a\partial _{\mu }\eta ^{a}\partial ^{\mu }
\chi ^{a}=-a\partial _{\mu
}\left( \eta ^{a}\partial ^{\mu }\chi ^{a}\right) 
+a\eta ^{a}\partial _{\mu
}\partial ^{\mu }\chi ^{a}\equiv a\eta ^{a}
\partial ^{2}\chi^{a}  \nonumber \\
&=&ag^{2}\eta ^{a}\chi ^{a}A_{\mu}^{b}A^{b\mu},
\label{(4.50)}
\end{eqnarray}
\begin{equation}
T_{11}+T_{15}=0,
\label{(4.51)}
\end{equation}
\begin{eqnarray}
T_{12} &=&g\varepsilon ^{abc}\partial _{\mu }
\left( \chi^{b}\eta^{c}\right)
\partial ^{\mu }\chi ^{a}=g\varepsilon ^{abc}\partial _{\mu }\left( \chi
^{b}\eta ^{c}\partial ^{\mu }\chi ^{a}\right) -g\varepsilon ^{abc}\chi
^{b}\eta ^{c}\partial _{\mu }\partial ^{\mu }\chi ^{a}\equiv -g\varepsilon
^{abc}\chi ^{b}\eta ^{c}\partial ^{2}\chi ^{a}  \nonumber \\
&&=-g^{3} {\underbrace{\varepsilon^{abc}\chi
^{a}\chi ^{b}}}\eta ^{c}A_{\mu }^{d}A^{d\mu},
\label{(4.52)}
\end{eqnarray}
i.e.
\begin{equation}
T_{12}=0,
\label{(4.53)}
\end{equation}
while
\begin{eqnarray}
T_{13} &=&g\partial _{\mu }\left( \chi ^{0}\eta ^{a}
\right) \partial^{\mu }\chi
^{a}\equiv -g\chi ^{0}\eta ^{a}\partial _{\mu }
\partial ^{\mu }\chi ^{a} 
\nonumber \\
&& =-g^{3}A_{\mu }^{b}A^{b\mu}
\eta^{a}\chi^{a}\chi^{0},
\label{(4.54)}
\end{eqnarray}
\begin{eqnarray}
T_{13}+T_{14}+T_{19} &=&
-g^{3}A_{\mu }^{b}A^{b\mu }\eta ^{a}\chi ^{a}\chi
^{0}-2g\partial _{\mu }\left( \chi ^{a}\eta ^{a}\right) 
\partial ^{\mu }\chi
^{0}+\frac{ag^{2}}{2}A_{\mu }^{b}A^{b\mu }\chi ^{a}
\eta ^{a}  \nonumber \\
&=&-g^{3}A_{\mu }^{b}A^{b\mu }\eta ^{a}\chi ^{a}\chi ^{0}
+2g\chi ^{a}\eta
^{a}\partial ^{2}\chi ^{0}
+\frac{ag^{2}}{2}A_{\mu }^{b}A^{b\mu }\chi
^{a}\eta ^{a}  \nonumber \\
&=&-g^{3}A_{\mu }^{b}A^{b\mu }\eta ^{a}\chi ^{a}
\chi ^{0}+g\chi ^{a}\eta
^{a}\left( g^{2}\chi ^{0}-\frac{ag}{2}\right) 
A_{\mu }^{b}A^{b\mu }+\frac{
ag^{2}}{2}A_{\mu }^{b}A^{b\mu }\chi ^{a}\eta ^{a}  \nonumber \\
&=&0.
\label{(4.55)}
\end{eqnarray}
Moreover
\begin{equation}
T_{17}+T_{18}=0,
\label{(4.56)}
\end{equation}
\begin{equation}
T_{20}=0,
\label{(4.57)}
\end{equation}
and we can exploit Eq. (4.25) to find
\begin{eqnarray}
T_{21} &=&-\frac{g}{2}\partial _{\mu }\left( b^{a}
\eta^{a}\right) \partial
^{\mu }e^{0}\equiv -\frac{g}{2}\left( b^{a}
\eta ^{a}\right) \partial _{\mu
}\partial ^{\mu }e^{0}  \nonumber \\
&& =-\frac{g^{3}}{2}A_{\mu }^{b}
A^{b\mu }b^{a}e^{0}\eta^{a},
\label{(4.58)}
\end{eqnarray}
\begin{equation}
T_{21}+T_{26}=0.
\label{(4.59)}
\end{equation}
By inspection, we also find the cancellation
\begin{equation}
T_{27}+T_{29}=0.
\label{(4.60)}
\end{equation}
Now we exploit Eq. (4.27) to find
\begin{eqnarray}
T_{22} &=&\frac{g}{2}\partial_{\mu}b^{0}\partial^{\mu}\left(e^{a}\eta
^{a}\right) \equiv -\frac{g}{2}\partial ^{\mu }\partial _{\mu }b^{0}\left(
e^{a}\eta ^{a}\right)   \nonumber \\
&& =-\frac{g^{3}}{2}A_{\mu}^{c}
A^{c\mu }b^{0}e^{a}\eta^{a},
\label{(4.61)}
\end{eqnarray}
\begin{eqnarray}
T_{25} &=&-\frac{g}{2}\partial ^{\mu}
\left( b^{0}\eta ^{a}\right) \partial
_{\mu }e^{a}\equiv \frac{g}{2}b^{0}\eta ^{a}
\partial^{\mu}\partial_{\mu}e^{a}  
\nonumber \\
&& =\frac{g^{3}}{2}A_{\mu}^{b}
A^{b\mu }b^{0}e^{a}\eta^{a},
\label{(4.62)}
\end{eqnarray}
\begin{equation}
T_{22}+T_{25}=0.
\label{(4.63)}
\end{equation}
At this stage, we can also exploit Eq. (4.26) to find 
\begin{eqnarray}
T_{23} &=&-\frac{g}{2}\varepsilon^{adc}
\partial ^{\mu }b^{a}\partial _{\mu
}\left( e^{d}\eta ^{c}\right) \equiv \frac{g}{2}\varepsilon ^{adc}\partial
_{\mu }\partial ^{\mu }b^{a}\left( e^{d}\eta ^{c}\right) \nonumber \\
&& =\frac{g^{3}}{2}\varepsilon^{adc}
A_{\mu}^{b}A^{b\mu}b^{a}e^{d}\eta^{c},
\label{(4.64)}
\end{eqnarray}
while Eq. (4.24) yields
\begin{eqnarray}
T_{24} &=&-\frac{g}{2}\varepsilon^{adc}
\partial ^{\mu }\left( b^{d}\eta
^{c}\right) \partial _{\mu }e^{a}\equiv \frac{g}{2}\varepsilon
^{adc}b^{d}\eta^{c}\partial^{\mu}\partial _{\mu}e^{a}  \nonumber \\
&=&\frac{g^{3}}{2}\varepsilon ^{adc}A_{\mu }^{b}A^{b\mu
}e^{a}b^{d}\eta ^{c} 
=-\frac{g^{3}}{2}\varepsilon^{adc}
A_{\mu}^{b}A^{b\mu}e^{d}b^{a}\eta^{c},
\label{(4.65)}
\end{eqnarray}
\begin{equation}
T_{23}+T_{24}=0.
\label{4.66)}
\end{equation}
By inspection, we also find the cancellations 
\begin{equation}
T_{27}+T_{29}=0,
\label{(4.67)}
\end{equation}
\begin{equation}
T_{28}+T_{30}=0.
\label{(4.68)}
\end{equation}
We are now left, in (4.38), with the following $5$ terms underlined 
here by a curly bracket and denoted hereafter by $A,B,C,D,E$:
\begin{eqnarray}
\delta L_{{\rm tot}} &=&
{\underbrace{g\varepsilon^{abc}F^{a\mu \nu
}\partial _{\mu }\left( A_{\nu }^{b}\eta ^{c}\right) }}
-{\underbrace{g\varepsilon^{abc}F^{a\mu \nu }A_{\mu }^{b}
\partial _{\nu }\eta
^{c}}}+{\underbrace{g^{2}F^{a\mu \nu }A_{\mu}^{b}A_{\nu
}^{a}\eta ^{b}}}  \nonumber \\
&&+{\underbrace{2g^{2}A^{a\mu}
\partial_{\mu}\eta^{a}\left(
\varphi ^{0}\varphi ^{0}+\chi ^{0}\chi ^{0}+b^{0}e^{0}+\varphi^{d}\varphi
^{d}+\chi ^{d}\chi^{d}+b^{d}e^{d}\right) }}  \nonumber \\
&&+{\underbrace{2agA^{a\mu}\partial_{\mu}\eta^{a}\left(
\varphi^{0}+\chi^{0}\right)}}.
\label{(4.69)}
\end{eqnarray}
Recall also that the equation of motion for the gauge field is
\begin{equation}
D_{\mu}F^{a\mu \nu }=2g^{2}A^{a\nu }
\left( \varphi ^{d}\varphi ^{d}+\chi
^{d}\chi ^{d}+b^{d}e^{d}+\varphi ^{0}\varphi ^{0}+\chi ^{0}\chi
^{0}+b^{0}e^{0}\right) +2agA^{a\nu }\left( \varphi ^{0}
+\chi^{0}\right),
\label{(4.70)} 
\end{equation}
or equivalently 
\begin{eqnarray}
\partial_{\mu }F^{a\mu \nu }+g\varepsilon^{abc}A_{\mu}^{b}F^{c\mu \nu }
&=&-2g^{2}A^{a\nu }\left( \varphi ^{d}\varphi ^{d}+\chi ^{d}\chi
^{d}+b^{d}e^{d}+\varphi ^{0}\varphi ^{0}+\chi ^{0}\chi
^{0}+b^{0}e^{0}\right)   \nonumber \\
&&-2agA^{a\nu}\left(\varphi^{0}+\chi^{0}\right).
\label{(4.71)} 
\end{eqnarray}
The first term in Eq. (4.69) then becomes
\begin{eqnarray}
A &=&g\varepsilon ^{aij}F^{a\mu \nu }\partial _{\mu }
\left( A_{\nu }^{i}\eta
^{j}\right) =g\varepsilon ^{aij}\left[ 
-\partial _{\mu }F^{a\mu \nu }\right]
A_{\nu }^{i}\eta ^{j}  \nonumber \\
&=&g\varepsilon ^{aij}\left[ g\varepsilon^{abc}
A_{\mu }^{b}F^{c\mu \nu
}+2g^{2}A^{a\nu }\left( \varphi ^{d}\varphi ^{d}+\chi ^{d}\chi
^{d}+b^{d}e^{d}+\varphi ^{0}\varphi ^{0}+\chi ^{0}\chi
^{0}+b^{0}e^{0}\right) \right]   \nonumber \\
&&+g\varepsilon ^{aij}\left[ 2agA^{a\nu }\left( \varphi ^{0}+\chi
^{0}\right) \right] A_{\nu }^{i}\eta^{j} \nonumber \\
&=&g^{2}\varepsilon ^{aij}\varepsilon ^{abc}F^{c\mu \nu }A_{\mu }^{b}A_{\nu
}^{i}\eta ^{j}  \nonumber \\
&&+2g^{3}{\underbrace{\varepsilon^{aij}
A^{a\nu}A_{\nu}^{i}}}
\eta ^{j}\left( \varphi ^{d}\varphi ^{d}+\chi ^{d}\chi
^{d}+b^{d}e^{d}+\varphi ^{0}\varphi ^{0}+\chi ^{0}\chi
^{0}+b^{0}e^{0}\right)  \nonumber \\
&&+2ag^{2} {\underbrace{\varepsilon^{aij}
A^{a\nu}A_{\nu}^{i}}}
\left(\varphi^{0}+\chi^{0}\right) \eta^{j}  \nonumber \\
&=&-g^{2}F^{a\mu \nu }A_{\mu }^{b}A_{\nu }^{a}\eta^{b}.
\label{(4.72)}
\end{eqnarray}
By inspection, the terms $A$ and $C$ cancel each other exactly: 
\begin{equation}
A+C=0,
\label{(4.73)}
\end{equation}
while the second term can be written as follows:
\begin{eqnarray}
B &=&-g\varepsilon ^{abc}F^{a\mu \nu }A_{\mu }^{b}\partial _{\nu }\eta
^{c}=\left( g\varepsilon ^{abc}A_{\mu }^{b}F^{c\mu \nu }\right) \partial
_{\nu}\eta^{a}  \nonumber \\
&&=-{\underbrace{\partial_{\mu }F^{a\mu \nu
}\partial _{\nu }\eta ^{a}}}-2g^{2}A^{a\nu }\partial _{\nu }\eta ^{a}\left(
\varphi ^{d}\varphi ^{d}+\chi ^{d}\chi ^{d}+b^{d}e^{d}+\varphi ^{0}\varphi
^{0}+\chi ^{0}\chi ^{0}+b^{0}e^{0}\right)   \nonumber \\
&&-2agA^{a\nu}\partial_{\nu}\eta^{a}\left( 
\varphi^{0}+\chi^{0}\right),
\label{(4.74)} 
\end{eqnarray}
and therefore 
\begin{equation}
B+D+E=0.
\label{(4.75)}
\end{equation}

Thus, we eventually obtain  
the important result that the new Lagrangian defined in
Eq. (3.4) is invariant under the shifted 
gauge transformations (1.9)--(1.17): 
$
\delta L_{{\rm tot}}=0.
$

\section{Concluding remarks and open problems}

Our paper does not contain new results but, having a purely
pedagogical character, aims at helping advanced readers to
become familiar with a new formalism for quantum Yang--Mills
theory as proposed in Ref. \cite{slav08}. In particular,
the calculations of Sec. IV, which prove in detail the
invariance of the S-matrix under shifted gauge transformations,
are worth reading for all those who are interested in modern
quantum field theory.

It now appears desirable to understand whether the Slavnov formalism
can be extended to quantum gravity, since Yang--Mills theory and
general relativity share the property of being type-I gauge theories
in a space-of-histories formulation \cite{dewi03,dewi08}. 
This means that the vector fields such that the action functional is
invariant under them have Lie brackets which are a linear combination
of the vector fields only with structure constants (i.e. independent
of the gauge fields). 

The counterpart of the Coulomb gauge considered in (1.1) by the
author of Ref. \cite{slav08} is the Prentki gauge, studied by
the authors of Ref. \cite{hoof74} in their pioneering work on 
one-loop divergences in quantum gravity. Such an extension to
quantum gravity would be, to our knowledge, original, and might
lead to a better understanding of this new class of gauge-fixed
functional integrals. 

\acknowledgments
H. Ghorbani thanks the Instituto de Fisica Te\'orica (IFT) in
Madrid for hospitality and support when this work was being done.
G. Esposito is grateful to the Dipartimento di Scienze Fisiche of
Federico II University, Naples, for hospitality and support; he
dedicates the present paper to Maria Gabriella.

\end{document}